\documentstyle[aps,preprint,epsfig]{revtex}
\begin{document}
\draft
\preprint{UFIFT-HEP-97-10}
\date{April 15, 1998}
\title{Caustic Rings of Dark Matter}

\author{P. Sikivie}
\address{Physics Department, University of Florida, Gainesville, FL
32611}

\maketitle

\begin{abstract}
It is shown that the infall of collisionless dark matter onto isolated 
galaxies produces a series of caustic rings in the halo dark matter 
distribution.  The properties of these caustics are investigated.  The
density profile of the caustic is derived for a specific case.  Bumps 
in the rotation curve of NGC 3198 are interpreted as due to caustic 
rings of dark matter.

\end{abstract}

\pacs{PACS numbers: 95.35.+d, 98.35.Gi}

\narrowtext

There are compelling reasons to believe that the dominant component of 
the dark matter of the universe is non-baryonic collisionless particles 
\cite{kt}.  The leading candidates are axions, WIMPs and massive neutrinos.  
The word ``collisionless'' signifies that the particles are so weakly 
interacting that they have moved purely under the influence 
of gravity since their decoupling at a very early time (of order 
$10^{-4}$ sec for axions, of order 1 sec for neutrinos and WIMPs).  In 
the limit where the primordial velocity dispersion of the particles is 
neglected, they all lie on the same 3-dim. 'sheet' in 6-dim. phase-space.  
Their phase-space evolution must obey Liouville's theorem.  This implies 
that the 3-dim. sheet cannot tear and hence that it satisfies certain 
topological constraints.

Let us assume that collisionless dark matter (CDM) exists.  Usually, CDM
means `cold dark matter' (e.g., axions or WIMPs) but because massive
neutrinos $(m_{\nu} >> 10^{-1}$ eV) behave, from the point of view of 
this paper, in a similar way we include them in our definition as well.  
Because their phase-space sheet cannot tear, CDM particles must be 
present everywhere in space, including specifically intergalactic 
space.  The space density may be reduced by stretching of the 
phase-space sheet but it cannot vanish.  Moreover, the average 
space density is recovered as soon as the average is taken over 
distances larger than the distance CDM may have locally moved 
away from perfect Hubble flow.  In a region which is sparsely 
populated with galaxies, this distance is much smaller than 
the distance between galaxies.  The implication is that isolated 
galaxies are surrounded by unseen CDM and hence, because of gravity, 
CDM keeps falling onto isolated galaxies continuously from all 
directions.  If the galaxy joins other galaxies to form a cluster, 
infall onto the galaxy gets shut off because of lack of material but 
infall onto the cluster continues assuming that the cluster is itself 
isolated.  In an open universe $(\Omega < 1)$, the infall process 
eventually turns off because the universe becomes very dilute.  However, 
even if our universe is open, we are far from having reached the 
turn-off time.  

It has been shown \cite{fg} that, under a wide range of circumstances, 
CDM infall onto an isolated galaxy produces a halo whose density falls
off like ${1\over r^2}$ for large $r$, where $r$ is the distance to the 
galactic center.  Such a halo implies a flat galactic rotation curve 
\cite{vcr}.  It has also been shown \cite{sty} that the angular momentum 
carried by the CDM particles has the effect of depleting the inner halo 
and of making the halo contribution to the rotation velocity vanish at 
$r\rightarrow 0$, thereby introducing an effective core radius.  Both 
the approximately flat rotation curves and the presence of effective core 
radii are consistent with observation.  CDM infall causes peaks in 
the velocity distribution of dark matter particles on Earth \cite{is,sty}.
The present paper focuses on the related phenomenon of ring caustics
in the halo density.  The existence of these caustics was noted in ref.
\cite{sty} but their properties were left unexplored.

Consider the infall of CDM onto an isolated galaxy.  Let's first
neglect the velocity dispersion of the infalling particles.  In practice
it is sufficient that the velocity dispersion of the infalling matter is 
much smaller than the rotation velocity of the galaxy but it is convenient 
to consider the extreme case of zero velocity dispersion first.  Consider 
the time evolution of all CDM particles which are about to fall onto 
the galaxy for the first time in their history at time $t$.  For the 
sake of definiteness, we may consider all particles which have zero 
radial velocity $(\dot{r} =0)$ for the first time then.  Such particles 
are said to be at their `first turnaround'; they were receding from 
the galaxy as part of the general Hubble flow before $t$ and will be 
falling onto the galaxy just after $t$.  They form a closed surface, 
enclosing the galaxy, hereafter called the turnaround 'sphere' 
at time $t$.  (The present turnaround sphere of the Milky Way 
galaxy has a radius of order 2 Mpc.)  The turnaround sphere at 
time $t$ falls through the central parts of the galaxy at a time 
of order $2t$.  Particles falling through the galactic disk 
(assuming the galaxy is a spiral) get scattered by an angle 
$\Delta \theta \sim 10^{-3}$ by the gravitational fields of various 
inhomogeneities such as molecular clouds, globular clusters and stars 
\cite{is}. However most particles carry too much angular momentum to 
reach the luminous parts of the galaxy and are scattered much less.  
Because the scattering is small, the particles on the turnaround sphere 
at time $t$, after falling through the galaxy, form a new sphere which 
reaches its maximum radius $R'$ at some time $t'$.  The radius $R'$ 
at the second turnaround is smaller than the radius $R$ at the first 
turnaround because the galaxy has grown by infall in the meantime.  
The sphere continues oscillating in this way although it gets 
progressively fuzzier because of scattering off inhomogeneities 
in the galaxy.

From a topological viewpoint, each time the sphere falls through the 
galaxy it turns itself inside out and hence there is a ring of special 
space-time points associated with each fall through the galaxy.  This 
ring may be defined as the locus of points which are inside the sphere 
last, just before the sphere completes the process of turning itself
inside out.  Fig. I illustrates this.  It shows successive 2-dim. 
cross-sections of a sphere falling in and out.  After the sphere has
crossed itself, as in frame (c), there is a donut-shaped volume of points
which are still inside the original sphere.  The ring is where this 
donut-shaped volume ends in space-time.  Because CDM falls in 
continuously, the ring in space-time associated with one infalling sphere 
is in fact a persistent feature in space.  For an arbitrary angular 
momentum distribution on the turnaround sphere, the ring is a closed loop 
of arbitrary shape.  However, if the angular momentum distribution is 
dominated by a smooth component which carries \emph{net} angular momentum, 
the ring resembles a circle.  If there is no angular momentum at all, the
ring reduces to a point at the galactic center.  As we will soon see, the 
ring is the location of a caustic with strong density contrast.  There is 
one caustic ring due to CDM particles falling through the galaxy for the 
first time, a caustic ring of smaller radius due to particles falling 
through for the second time, a yet smaller ring due to particles falling 
through for the third time, and so on.  The caustic rings move slowly in 
space, generally in an outward direction, as the properties of the infalling 
CDM particles (in particular, their turnaround radius and angular momentum 
distribution) change.

Let's obtain the density profile of the caustic in a particular case of
axial symmetry about the z-axis and parity (z$\rightarrow$-z).  The
symmetry assumptions will force the caustic to lie on a circle in the
z=0 plane.  Consider the time evolution of CDM particles initially located 
with uniform density on a turn-around sphere of radius $R$ and with 
the angular momentum per unit mass distribution 
$\vec\ell(\hat r) = \omega \vec r \times (\hat z \times \vec r)$ 
where $\vec r = R\hat r$.  Thus the turnaround sphere is assumed to be 
initially rotating about the z-axis with angular velocity $\omega$ 
as if it were a rigid body.  Let's assume further that the particles 
fall into the spherically symmetric potential:  $V(r) = v_{rot}^2 \ln (R/r)$, 
which is such that it produces a flat rotation curve with rotation
velocity $v_{rot}$.  One readily finds that a continuous flow of
particles falling through under the stated conditions produces a density
distribution which has a caustic ring in the $z=0$ plane of radius $a$
given by:
\begin{equation}
a~v_{rot} \sqrt{2 \ln (R/a)} = w R^2 (1+0({a^2 \over R^2}))\ .
\label{(1)}
\end{equation}
The caustic is the locus of points where the particles 
with the largest amount of angular momentum $(\ell_{max} = \omega R^2)$ 
are at their distance of closest approach to the galactic center.  A
calculation shows that, to leading order when $a<<R$, the density 
distribution is given by:
\begin{equation}
d(a;\rho,z) = {dM\over d\Omega dt}~ {2\over v}~ {1\over
\sqrt{(r^2-a^2)^2 + 4a^2 z^2}}
\label{(2)}
\end{equation}
where $(\rho,z,\theta)$ are cylindrical coordinates, $r = \sqrt{\rho^2 +
z^2},~ {dM\over d\Omega dt}$ is the rate at which mass falls in per unit
time and unit solid angle, and $v = v_{rot} \sqrt{2 \ln (R/a)}$ is the
velocity of particles near $r=a$.  From Eq. (2), one readily finds the
behaviour of the density near the caustic:
\begin{equation}
d(a;\rho,z) \simeq {dM\over d\Omega dt}~ {1\over v~ a~ \sigma}
\label{(3)}
\end{equation}
where $\sigma = \sqrt{(\rho -a)^2 + z^2}$ is the distance to the ring.
Of course the caustic singularity is only perfectly sharp in the limit 
where the velocity dispersion of the infalling CDM is zero.  If the 
velocity dispersion is $\Delta v$, the caustic singularity spreads over 
a distance of order $\Delta a = R {\Delta v\over v}$.

For an isolated galaxy, there is a ring of radius $a_1$ due to particles 
falling through for the first time, a ring of radius $a_2$ due to 
particles falling through for the second time, and so on.  Thus the 
quantities $R,~a,~v,~d$ and ${dM\over d\Omega dt}$ carry an index 
$n = 1,2,3...$.  One caustic ring is associated with each pair of 
velocity peaks \cite{is,sty} in the halo distribution.  Eq. (2) 
implies ${dM_n\over d\Omega dt} {2\over v_n} = r^2 d_n (0;r)$.  This 
relates the value of the prefactor
${dM_n\over d\Omega dt} {1\over v_n}$ for each infall to the density 
$d_n(0;r)$ the associated pair of velocity peaks contributes in the 
limit of zero angular momentum ($a=0$).  Estimates of $d_n(0,r)$ 
can be found in ref.\cite{sty} for the case of self-similar infall.  

The infall is called self-similar \cite{fg,bert} if it is time-independent 
after all distances are rescaled by the turn-around radius $R(t)$ 
at time $t$ and all masses are rescaled by the mass $M(t)$ interior 
to $R(t)$.  In the case of zero angular momentum and spherical symmetry, 
the infall is self-similar if the initial overdensity profile has the form 
${\delta M_i\over M_i} = ({M_0\over M_i})^\epsilon$ where $M_0$ and 
$\epsilon$ are parameters \cite{fg}.  $\epsilon$ must be in the range 
$0 \leq \epsilon \leq 1$.  The rotation curve is flat if 
$0 \leq \epsilon \leq 2/3$ \cite{fg}.  The infall model was generalized in
ref. \cite{sty} to include the effect of angular momentum. It was found that 
self-similarity requires the angular momentum distribution $\ell(t)$ to 
have the time-dependence $\ell (t) = j {R^2(t)\over t}$, where $j$ is a 
dimensionless and time-independent distribution. A good agreement of the 
self-similar model with the properties of our own galaxy was found 
\cite{sty} for parameter values of order $\epsilon = 0.25,
~\bar j = 0.2$ and $h = 0.7$ where $\bar j$ is the average of the $j$ 
distribution and $h$ is the Hubble rate in units of 100 km/sec.Mpc .  
Using the model of ref.\cite{sty}, the following values for the radii of 
the first five rings are obtained:
\begin{equation}
\{a_n: n=1,2...5\} \simeq (39, 19.5, 13, 10, 8) kpc \left({j_{max}\over
0.25}\right) \left({0.7\over h}\right) \left({v_{rot}
\over 220 {km\over s}}\right)
\label{(4)}
\end{equation}
where $j_{max}$ is the maximum $j$ value of the $j$ distribution, and
where the value $\epsilon = 0.3$ was used.  For $\epsilon = 0.2,~a_1
\simeq 36 kpc~\left({j_{max}\over 0.25}\right) \left({0.7\over h}\right) 
\left({v_{rot}\over 220 km/s}\right)$ but the ratios $a_n/a_1$ are almost 
the same as in the $\epsilon = 0.3$ case. 

Luminous rings surrounding galaxies have been observed and such rings 
may be related to the caustics described here.  However, luminous matter 
is presumably baryonic and the behaviour of baryons is more complicated 
than that of CDM particles.  Infalling baryons do not necessarily behave 
in a collisionless manner and they may easily get stripped off the 
CDM flow.  In contrast, the conclusion that isolated galaxies are 
surrounded by caustic rings of \emph {dark} matter appears unavoidable 
if CDM exists.  Because of this, we will limit ourselves in this paper 
to the observational implications of CDM rings which follow from their 
gravitational fields.  CDM rings may nonetheless have a baryonic 
component.  These baryons may have the same phase-space distribution as 
the CDM or they may have accreted onto the ring from neighboring space. 

The Newtonian gravitational force per unit mass due to the density 
distribution of Eq. (3) is
\begin{equation}
\vec F_n (\vec \sigma) = - {2\pi G~C_n\over a_n} \hat{\sigma} 
\label{(5)}
\end{equation}
for small $\sigma$, where $C_n \equiv {dM_n\over d\Omega dt}
{2\over v_n} = r^2~d_n (0;r)$.  Hence there is a discontinuity in 
the rotation velocity
\begin{equation}
{\Delta_n v_{rot}\over v_{rot}} = {1\over 2} {d_n (0;a_n)\over
d(a_n)} ~~~~~,
\label{(6)}
\end{equation}
directly across the $n^{th}$ caustic, in the ideal case where this
caustic is perfectly sharp and lies in the galactic plane.  In Eq. (6), 
$d(r) \equiv {v_{rot}^2\over 4\pi Gr^2}$ is the total density at $r$ 
in the limit of spherical symmetry and perfectly flat rotation curves.  
The ratios $f_n \equiv {1\over 2} {d_n (0;r)\over d(r)}$ were found 
to be of order (13, 5.5, 3.5, 2.5, 2, ...)$10^{-2}$ in the self-similar 
infall model with $\epsilon = 0.2$.  (For $r << R$ and small n, the 
$f_n$ are nearly $r$-independent.) There are of course a number of 
effects that will smooth out sudden variations in the measured rotation 
velocities.  One effect is that the CDM ring may be some distance away 
from the galactic plane where the rotation velocities are measured.  
Secondly,  the measured rotation velocities are spatial averages over 
some distance across the galactic plane.  Thirdly, the CDM ring may  
be fuzzy because of velocity dispersion as mentioned earlier.  Thus 
the discontinuities of the ideal case are smoothed out into bumps.  
The bumps occur in the measured rotation curve near the location of 
CDM rings if the latter happen to be close to the galactic plane.

Galactic rotation curves often do have bumps.  Of special interest here
are those bumps which occur at radii larger than the disc radius because
they cannot readily be attributed to inhomogeneities in the luminous
matter distribution.  Consider the rotation curve of NGC 3198 \cite{vA},  
one of the best measured and often cited as providing compelling 
evidence for the existence of dark halos.  It appears to have
bumps near 28, 13.5 and 9 kpc, assuming $h = 0.75$.  Although the
statistical significance of these bumps is not great, let's
assume for the moment that they are real effects.  Note then that their
existence is inconsistent with the assumption that the dark halo is 
a perfect isothermal sphere.  On the other hand, the radii at which 
the bumps occur are in close agreement with the ratios:
\begin{equation}
\{a_n/a_{n-1}: n=2,3,4,5\} = (0.49,~0.67,~0.76,~0.81)
\label{(7)}
\end{equation}
predicted by Eq. (4) assuming that the bumps are caused by the 
gravitational fields of the first three caustic rings of NGC 3198.  
Since $v_{rot} = 150$ km/sec, we find that $j_{max} = 0.28$ in this case
if $\epsilon = 0.3$.  The uncertainty in $h$ drops out.  A fit of the 
infall model to our own galaxy \cite{sty} produced $\bar j \simeq 0.2$ for
$\epsilon = 0.2$ to $0.3$. If the turnaround sphere is taken to be rigidly 
rotating, one has $j_{max} = {4\over \pi}\bar j$.  Thus the values of 
$\bar j$ for our own halo and that of NGC 3198 are found to be similar. 

The ratios of caustic ring radii given in Eq.(7) are characteristic
of the $t$-dependence in the angular momentum distribution 
$l(\hat r,t) = j(\hat r) R(t)^2/t$ .  The main reason for using this 
ansatz in ref. \cite{sty} was that it produces exact self-similarity in 
the time evolution of the halo.  However, there is a broader justification 
for a time-dependence close to the one of the ansatz.  Consider particles 
which at some early initial time $t_i$ are at a distance $r_i$ from the 
center of a large overdensity which will grow into a galactic halo.  These 
particles have magnitude of angular momentum with respect to the 
overdensity's center: $l(\vec r_i) = r_i v_{i\perp} (\vec r_i)$ where 
$v_{i\perp}$ is the magnitude of the component of the initial velocity 
$\vec v_i(\vec r_i)$ transverse to $\vec r_i$.  Because 
$\vec v (\vec 0) = 0$, it is reasonable to assume $v_{i\perp}(\vec r_i)$ 
is proportional to $r_i$ when comparing values of $r_i$ which are 
of the same order of magnitude.  In that case,
$l(\vec r_i) \sim r_i^2 \sim M_i^{2 \over 3} \sim 
R(r_i)^2/t(r_i)^{4 \over 3}$ where $M_i$ is the mass interior at time
$t_i$ to the sphere of radius $r_i$, $R(r_i)$ and $t(r_i)$ are the 
turn-around radius and turn-around time of particles initially at
radius $r_i$, and where we used $G M_i = \pi^2 R(r_i)^3/8 t(r_i)^2$.
Except for the relatively slowly-varying factor of $1/t^{1\over 3}$, 
this is the time dependence which leads to self-similarity.  
At any rate, since the gravitational field around a large isolated 
overdensity is approximately spherically symmetric and hence angular 
momentum about the center of the overdensity is approximately conserved, 
the study of bumps in rotation curves may inform us about the peculiar 
velocities associated with primordial density perturbations and thus 
constrain theories on the origin of these perturbations.

Finally, let's discuss the implications for our own halo of the 
assumption that the caustic rings lie in the galactic plane.  As
an example, we will use the model parameters $\epsilon = 0.25$, 
$j_{max} = 0.265$ and $h=0.7$. In this case the Milky Way caustic rings 
are at the radii:  41, 20, 13, 10, 8.0, 6.7, 5.8, 5.1, 4.5, 4.0, 3.7, 
3.4, 3.1 ... kpc.  We at 8.5 kpc would be between the 4th and 5th ring.  
There is evidence for the 6th through 13th caustic rings in that there
are sudden rises in the inner rotation curve published in ref.\cite{Cle}
at radii very near those listed above. Similarly there is some evidence 
for the 2d and 3d caustic rings in the averaged outer rotation 
curve published in ref.\cite{Tre}.  This will be discussed in detail in 
a future paper\cite{tbp}.  The 4th ring would have passed by us 
approximately 900 million years ago assuming that our distance to the 
galactic center is and has always been 8.5 kpc.  The 5th ring would
pass by us approximately 350 million years from now.  It is conceivable
that the passage of caustic rings caused cataclysmic events in our past.
If the angular momentum distribution is the one 
${dn\over dj} = {j\theta (j_{max} - j)\over j_{max} (j_{max}^2 -j^2)}$ 
characterizing a turn-around sphere which is initially rigidly rotating,
the contributions of the first eight incoming peaks to the local halo 
density would be approximately (0.4, 0.9, 1.7, 5, 13, 3, 2, 1.4)
$10^{-26} {gr\over cm^3}$ whereas the averages \cite{sty} over all 
locations at the same distance from the galactic center as us are 
approximately (0.4, 0.8, 1.2, 2, 2.7, 1.6, 1.2, 1.0) $10^{-26} 
{gr\over cm^3}$.  The 4th and 5th velocity peaks are considerably 
enhanced because of our proximity to the corresponding rings.  In the 
model, the two flows associated with the 4th ring have velocity vectors 
$(210 \hat \phi \pm 260 \hat z)$km/s in the Sun's rest frame whereas 
those associated with the 5th ring have velocity vectors 
$(230 \hat \phi \pm 160 \hat r)$km/s.  Here, $\hat z$ is the direction 
perpendicular to the galactic plane, $\hat r$ is the radial direction 
in the galactic plane and $\hat \phi$ is the direction of rotation in 
the plane.  If the caustic rings lie in the plane, the local halo 
density is boosted from approximately $0.5~10^{-24} {gr\over cm^3}$, 
which is the usual estimate in the case of spherical symmetry, to 
approximately $0.9~10^{-24} {gr\over cm^3}$.  These results are 
relevant to the axion \cite{ad} and WIMP \cite{sl} dark matter searches 
which in fact provided the original impetus for this work.  It is 
conceivable that these experiments will measure some day the 
contributions of the velocity peaks to the local density and thereby 
provide us with detailed information about the structure of our halo.
   
\acknowledgments
I am very grateful to I. Tkachev for the use of his self-similar infall 
computer programs and to D. Eardly and J. Ipser for stimulating comments.  
This work was supported in part by a grant from the J.S. Guggenheim 
Memorial Foundation and by DOE grant DE-FG05-86ER40272.

\begin{figure}
\vspace{1.5cm}
\epsfxsize=6in
\centerline{\epsfbox{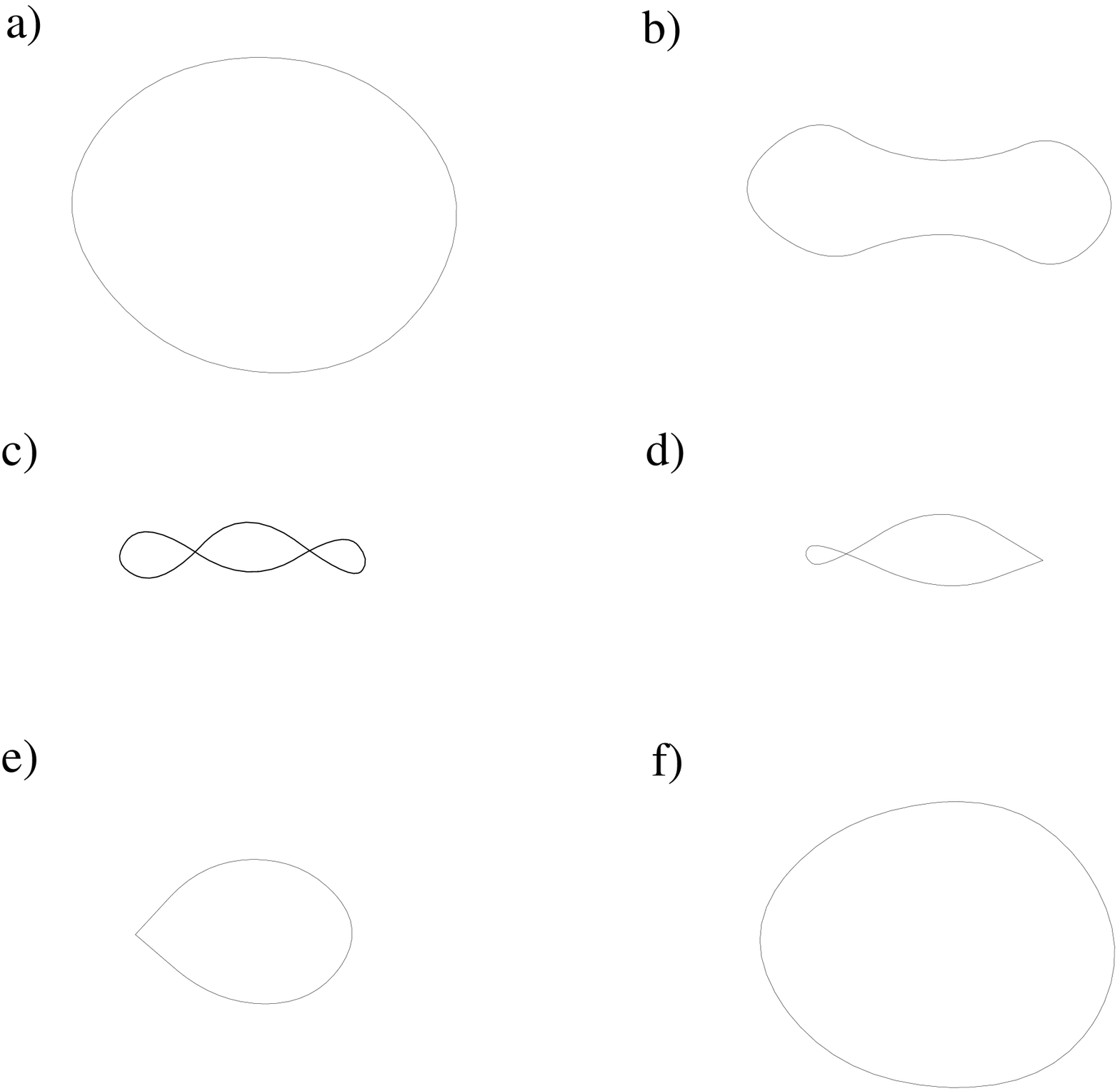}}
\vspace{2.5cm}
\caption{Infall of all particles on a turnaround sphere.  The sphere 
has net angular momentum about the vertical axis.  It crosses 
itself between frames b) and c).  The cusps in frames d) and e) are 
at the intersection of the ring caustic with the plane of the figure.}
\label{fig:infall}
\end{figure}

\end{document}